\documentclass{ws-p9-75x6-50}
\usepackage{epsfig}
\def\figwidth{8cm}

\begin{document}

\title{Interactions in Quasicrystals}

\author{Julien Vidal}
\address{Laboratoire de physique des solides, CNRS UMR 8502,
UPS B{\^a}t. 510, 91405 Orsay, France \\ Email:
vidal@lps.u-psud.fr }

\author{Dominique Mouhanna}
\address{Laboratoire de Physique Th\'{e}orique et
Hautes Energies, CNRS UMR 7589, Universit\'{e} Pierre et Marie
Curie Paris 6, \\ 4 place Jussieu, 75252 Paris Cedex 05 France
\\ Email: mouhanna@lpthe.jussieu.fr}

\author{Thierry Giamarchi}
\address{Laboratoire de physique des solides, CNRS UMR 8502,
UPS B{\^a}t. 510, 91405 Orsay, France \\ Email:
giam@lps.u-psud.fr}


\maketitle

\abstracts{Although the effects of interactions in solid state
systems still remains a widely open subject, some limiting cases
such as the three dimensional Fermi liquid or the one-dimensional
Luttinger liquid are by now well understood when one is dealing
with interacting electrons in {\it periodic} crystalline
structures. This problem is much more fascinating
 when periodicity is lacking as it is the case in {\it
quasicrystalline} structures. Here, we discuss the influence of the
interactions in quasicrystals and show, on a controlled
one-dimensional model, that they lead to anomalous transport
properties, intermediate between those of an interacting
electron gas in a periodic and in a disordered potential.}

\section{Introduction}

Our progress in solid state physics have been drastically
dependent on our understanding of the physical properties of
periodic structures. The knowledge of the eigenstates (Bloch
waves) in such structures, enabled us to tackle other physical
effects such as electron-electron interactions and
transport properties. Some studies have also been made to understand systems
totally lacking periodicity such as disordered ones. In this
case also the knowledge of the diffusive, or exponentially
localized nature of the wavefunctions has been a good starting point. An
intriguing situation occurs in the case of quasicrystals
\cite{Shechtman}, for which the system is neither periodic nor
disordered. On the experimental side the transport exhibits
many unusual properties. These metallic alloys are notably characterized by a low
electrical conductivity $\sigma$ which increases when either
temperature or disorder increases \cite{Mayou93}. The very low
temperature behaviour of $\sigma$ is still an open question and
strongly depends on the materials. For example, in $AlCuFe$ \cite{Klein}
and $AlCuRu$ \cite{Biggs}, a finite conductivity at zero
temperature is expected whereas recent results \cite{Delahaye}
seem to confirm a Mott's variable range hopping mechanism
($\sigma(T)\sim \exp-(T_0/T)^{1/4}$) for $i-AlPdRe$ icosahedral
phase down to $20$ mK. The optical conductivity is also unusual
since there is no Drude Peak for icosahedral quasicrystals
\cite{Homes,Burkov}.

 From a theoretical point of view, the case of independent
electrons in one-dimensional ($1D$) systems has been deeply
investigated for different quasiperiodic structures (Harper model,
Fibonacci chain,..)\cite{KKT,Ostlund}, giving rise to singular
continuous spectra with an infinite number of gaps. Moreover, the
corresponding eigenstates are neither extended nor localized but
critical, and are known to be responsible of anomalous diffusion
\cite{Zhong,Piechon}. For higher dimensional systems (Penrose
tiling, Octagonal tiling, icosahedral structure...), similar
studies had also displayed complex and intricated spectra, with
analogous characteristics of the electronic states
\cite{Passaro_Octo_diffusion,Yamamoto_Penrose}. Finally, studies
of transport
properties\cite{Mayou99,Sire_Aussois,Roche_Review,Roche_Fermi,Bellissard_aperiodic}
has also displayed anomalous behaviour such as a power-law
decreasing of the conductance with the system size
\cite{Tsunetsugu_conductance,Kubo_Goda}.

Given the difficulty of the full quasiperiodic problem, and the
relatively poor knowledge of the single particle wavefunctions,
adding interactions is a very difficult task. Even in $1D$
incommensurate structures, few results have been obtained
\cite{Hiramoto_HF,Chaves_meso,Chaves_HF,Mastro,Sen}. In order to
treat interactions in quasicrystals we thus follow a different
route, used with success for disordered systems. We first solve
the {\it periodic} system in presence of interactions. This is
relatively easy, either in the one-dimensional case which is
considered in this paper, for which exact solutions exists, or
even in higher dimensions through approximate (Fermi liquid)
solutions. Then, we study the effect of a weak perturbative
quasiperiodic potential on this solution via a renormalization
group approach. The main idea consists in analyzing the dependence
of the low-energy physics with respect to the Fourier spectrum of
the potential. The influence of the renormalization group
procedure is discussed in terms of an effective elimination of the
various potential harmonics around the Fermi level, leading to
unusual properties in the case of a self-similar quasiperiodic
structure\cite{vidal_boso}.

\section{The model}
Let us consider a system of interacting spinless fermions on a
lattice embedded in a general on-site (diagonal) potential $W$, described by the
following Hamiltonian:
\begin{eqnarray}
H&=&-t \sum_{\langle i,j \rangle} c^\dagger_{i}\,c_{j}+V \sum_{i}
 n_i\,n_{i+1}
+ \sum_{i} W_i\,n_i\\
&=&H_0(t,V)+H_W
\label{hamil}
\end{eqnarray}
where $c^\dagger_{i}$ (resp. $c_{i}$) denotes the creation (resp.
annihilation) fermion operator,
 $n_i=c^\dagger_{i}\,c_{i}$ represents the fermion density on site $i$, and
$\langle \ldots \rangle$ stands for nearest neighbors pairs.

In the continuum limit, $H_0(t,V)$ writes\cite{schulz_houches_revue}:
\begin{equation}
H_0(t,V)={1\over 2\pi}\int dx \left[(u K)(\pi \Pi)^2+\left({u\over K}\right)
(\partial_x \phi)^2\right]
\label{H0bos}
\end{equation}
where $\phi$ is a boson field related to the long wave length part
of the fermionic density by $\rho(x)=-\nabla \phi(x)/\pi$, and
$\Pi$ is its canonically conjugate field. This approximation provides
the correct description of the low energy physics.
All the interactions are absorbed in the two constants $u$ and $K$. $u$ is the renormalized Fermi
velocity (in the non interacting case: $u=v_F=2ta\sin(k_F a)$).
K is the Luttinger parameter controlling the decay of various correlation functions:
K=1 in the non-interacting case, $K>1$ for attractive interactions $(V<0)$
and $K<1$ for repulsive interactions $(V>0)$.
Note that, for many models, these parameters can be  explicitely  computed
in terms of the microscopic details \cite{haldane_bosonisation,haldane_xxzchain}.

The perturbative part of $H$, can also be expressed in terms of
these boson fields:
\begin{equation} \label{eq:general}
H_W = \frac1{\pi \alpha} \int dx \,W(x)
\cos\left[2k_Fx-2\phi(x)\right] \label{hquasi}
\end{equation}
where $\alpha$ is a short distance
cut-off of the order of the lattice constant $a$.

\section{Renormalization group analysis}

To determine the long distance physics, we use a standard
perturbative renormalization group approach. Full details can be
found in \cite{vidal_boso,vidal_quasiinter_long} and we focuss
here on the main equations for the potential and interaction
parameter:
\begin{eqnarray}
{dK\over dl}&=&-{K^2 \over 2} G(l)  \nonumber \\
{dy_q\over dl}&=&(2-K) y_q\label{eq:rgeq}
\end{eqnarray}
with
\begin{equation}
G(l) =\sum_{\varepsilon=\pm 1}\sum_q y_q^2 J\left[(q+\varepsilon
2k_F)\,\alpha(l) \right]\ .
\label{recy3}
\end{equation}
$y_q=\alpha \hat W(q) / u$ is the dimensionless Fourier
components of $W$ and  $\alpha(l)=\alpha(0) \,e^l$ is the renormalized
short distance cutoff. In Eq. (\ref{recy3}), $J$ is an ultraviolet
regulator whose precise form depends on the procedure used to
eliminate the high energy degrees of freedom. Typically one has:
\begin{eqnarray}
J(x)&\simeq& 1 \hspace{0.5cm} {\hbox{if}} \hspace{0.5cm} x<1\\
&=& 0\hspace{0.5cm}  {\hbox{otherwise}}
\end{eqnarray}

Physically, this renormalization is equivalent to an investigation
of the low-energy properties in a window around $2k_F$ in the
reciprocal space whose width is proportionnal to $e^{-l}$. Thus,
the {\it full} Fourier landscape of the potential determine the
scaling of the parameters $u$ and $K$, and one has to distinguish
several situations.

\begin{figure}[p]
 \centerline{\psfig{file=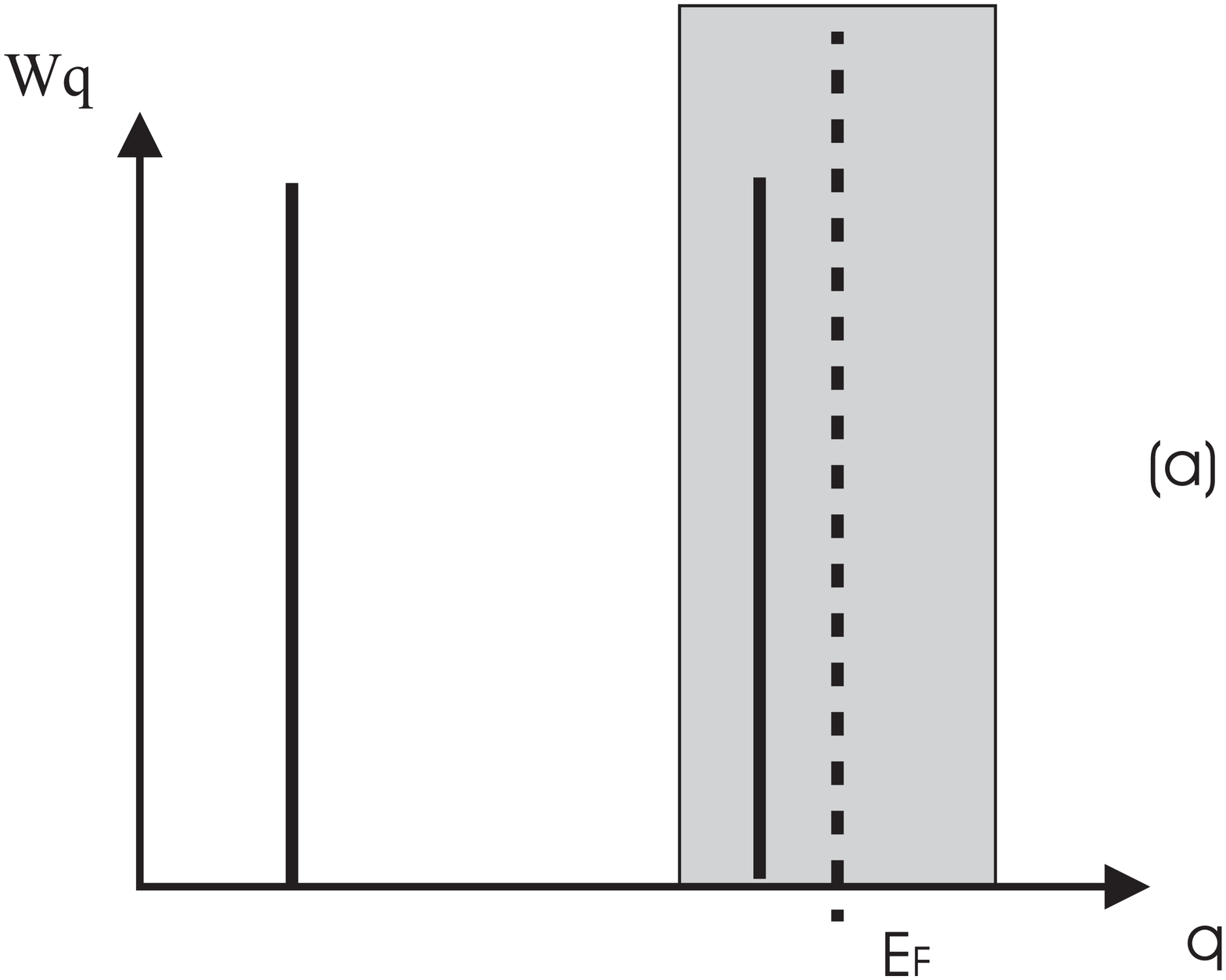,angle=0,width=\figwidth}}
 \centerline{\psfig{file=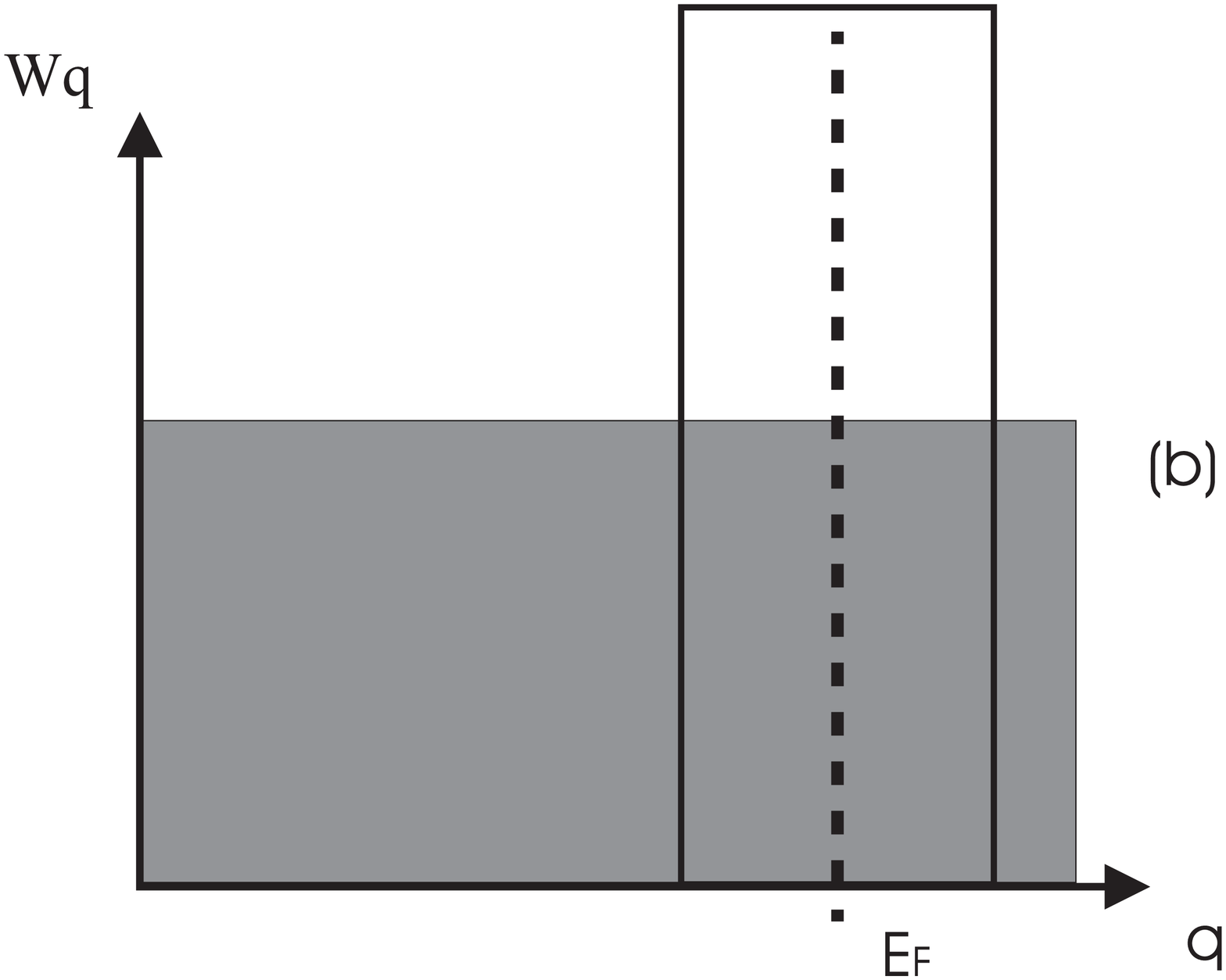,angle=0,width=\figwidth}}
 \centerline{\psfig{file=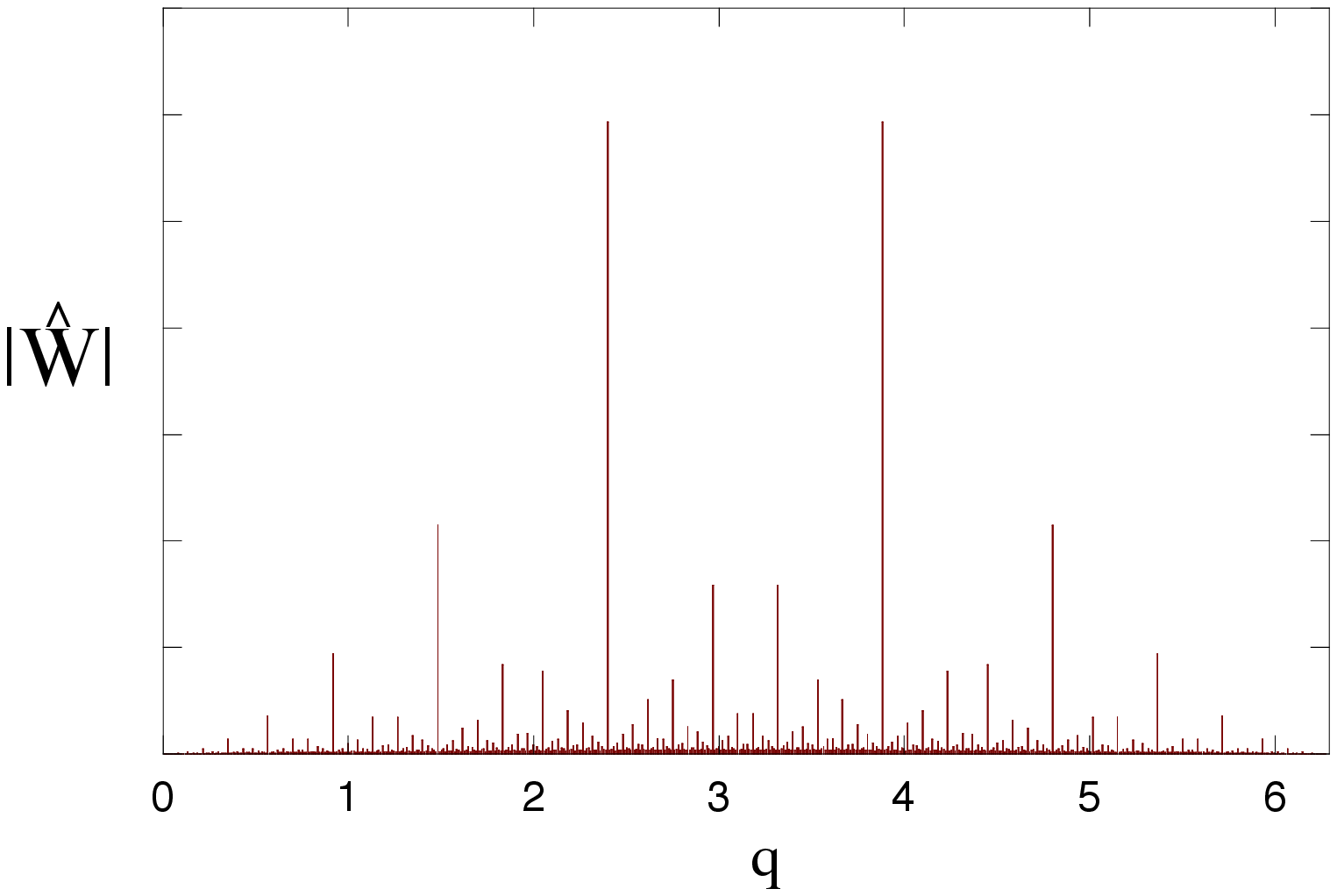,angle=0,width=\figwidth}}
  \caption{\label{fig:fourier} Fourier spectrum of the potential $W$ for the
  periodic (a), disordered (b) or quasiperdiodic (c) cases. All the peaks within a window of
  width $\sim e^{-l}$ (see text) around the Fermi energy $E_F$ (shown as a box) control
  the physics of the problem. This is crucial for the disordered (b) and quasiperiodic (c) systems for
  which it is impossible to really isolate a single peak, contrarily to the case of periodic
  systems (a)}
\end{figure}

If the Fourier spectrum of $W$ has only a single peak or few well
separated peaks, then the system is analogous to a periodic
system. If the Fermi level is not on one of the peaks then at some
lengthscale the regulator $J$ gives always zero (see
figure~\ref{fig:fourier}). The potential $W$ is irrelevant and one
recovers a Luttinger liquid metallic phase. On the contrary, if
the Fermi level is right on one of the peaks (e.g. $2 k_F=q_0$),
all physical properties are dominated by this single harmonic (the
regulator cuts all the others) and the system feel a periodic
potential $W(x)= \lambda \cos(q_0x)$. The case of the single
harmonic is well known
\cite{luther_exact,giamarchi_umklapp_1d,kolomeisky_periodic_diagphas}:
the periodic potential is relevant for $K<2$ and opens a gap
\begin{equation}
\Delta\sim y_q^{1\over 2-K}
\end{equation}
For $K>2$ the periodic potential is irrelevant and system remains
a Luttinger liquid with gapless excitations.
Our equations clearly show that novel physics occurs
when the spectrum of the potential $W$ is dense. This is the case
both for disordered systems and for the quasiperiodic ones
(see figure~\ref{fig:fourier}). Indeed in that case it is never
possible to really isolate a single peak \cite{footnote_harper}.
The physics is thus {\it not} controlled by the presence or not of a peak {\it at} the
Fermi level but by the forest of peaks in the (shrinking) window
of width $e^{-l}$. In this respect, the properties of a
quasicrystal are  much closer (with proper differences discussed
below) to the one of a disordered system than the one of a
``periodic'' (i.e. with a limited number of harmonics) one.

The disordered case can easily be obtained from our equations.
Let us assume that $W$ is uncorrelated with averages given by
\begin{equation}
\overline{\hat W^*(q)\hat W(q')}=\lambda^2\,\delta_{qq'}
\end{equation}
In the limit of weak disorder, (\ref{eq:rgeq}) can be integrated
neglecting the renormalization of $K$: $y(l)=y(0)\,e^{(2-K)l}$.
Then, (\ref{eq:rgeq}) simply becomes:
\begin{equation} \label{eq:mu}
\frac{dK}{dl} = - K^2 C \,e^{(3-2K)l}
\end{equation}
where $C$ is a constant. Eq. (\ref{eq:mu}) provides a critical
value $K_c=3/2$, separating an Anderson-like insulating phase
($K<K_c$) from a metallic state ($K>K_c$)
\cite{giamarchi_loc,apel}. Note that contrary to the periodic
case, this metal-insulator transition occurs for any  value of
$k_F$. Indeed, as shown figure~\ref{fig:fourier}, the Fourier
spectrum is uniform (after averaging) so that all the positions of
the Fermi level are equivalent.

\section{The quasiperiodic case}

The quasiperiodic case provides an interesting intermediate
situation. Let us consider for simplicity a
Fibonacci potential although our results can be easily extended to
more general situations.

The quasiperiodicity is provided by the $W_i$'s
that take two discrete values $W_A=+\lambda/2$ or $W_B=-\lambda/2$
given by the spatial modulation of the Fibonacci chain. In
fact, we consider a periodic approximant
of this structure with $F_j$ sites per unit cell that can be obtained
 by $j$ iterations of the substitution rules: $A\rightarrow
AB,\hspace{2.ex} B\rightarrow A$,
where $F_j$ is the $j^{th}$ element of the Fibonacci sequence defined by:
\begin{eqnarray}
F_1&=&F_2=1 \nonumber\\
F_{j+1}&=&F_{j}+F_{j-1}
\label{recur}
\end{eqnarray}
We denote $p=F_{j-2}, s=F_{j-1}, n=F_{j}$ and $n'=s$ (resp. $n'=p$) if
$j$ is even (resp. odd). In the quasiperiodic limit ($j\rightarrow
\infty$), the ratio $s/p$ converges toward the golden mean
$\tau={1+\sqrt{5} \over 2}$.
The Fourier transform of  $W$ can be straightforwardly
using the conumbering scheme \cite{Mosseri_conumbering}:
\begin{equation}
\hat W\left(q={2\pi m\over na}\right)={\lambda \, e^ {i{\pi m  n' (s-1)
\over n}} \sin
\left({\pi m n' s \over n}\right)
\over n \sin \left({\pi m n'\over n}\right)}
\label{TF}
\end{equation}
for $m=1$ to $n-1$ ($a$ is the lattice spacing). A global shift of
the $W_i$ allows us to deal with a zero-averaged potential so that
we can set $\hat W(0)=0$. Moreover, the underlying substitution
rule provides a self-similar structure that can be readily seen in
Fig.~\ref{fig:fourier}. In the non interacting case, each Fourier
component of the potential opens a gap  whose width is given, at
first order in perturbation, by the amplitude of the corresponding
component\cite{Sire_pertu}. In the quasiperiodic limit, the
Fourier spectrum of the potential become dense so that the
spectral measure goes to zero. Therefore, for any filling factor,
the Fermi velocity vanishes and the system is  an insulator (at
zero temperature). In contrast, given the complexity of the
Fourier spectrum, one can expect the long distance physics of the
interacting system to depend on the position of the Fermi level.
Indeed, the flow equations (\ref{recy3}) show that the behaviour
of $G$ is strongly sensitive to the position of the window in the
spectrum. For a given maximum renormalization length scale
$l_{max}$,corresponding to accessible physical range, two
different cases must be distinguished.

First, if the Fermi momentum $2k_F$ is close to a main peak of the
Fourier spectrum, {\it i.\,e.} $(q\pm 2k_F)^{-1}>l_{max}$ then, at
long distance, i.e. up to $l\sim l_{max}$, the flow of $K$ is
controlled by this harmonic  (see Fig.~\ref{fig:fourier} (upper
curve)). A metal-insulator transition can occurs with $K_c=2$.
This case is very similar to the simple periodic one.

\begin{figure}[h]
 \centerline{\psfig{file=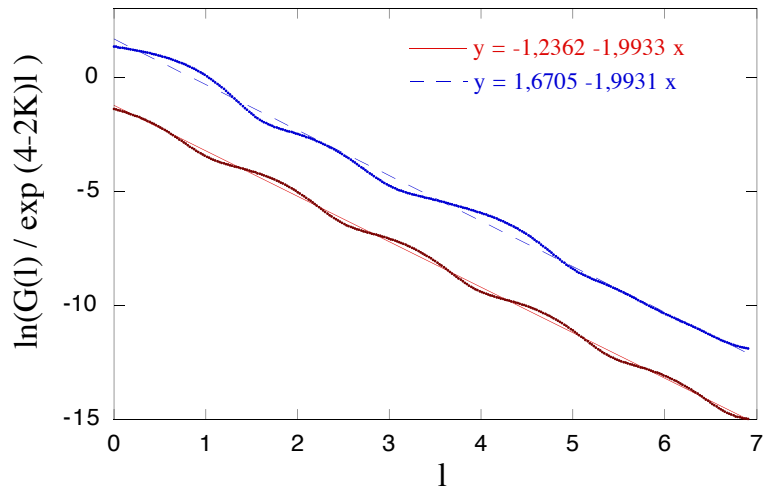,angle=0,width=\textwidth}}
  \caption{\label{fig:smallpeak} Scaling of the function $G$. The presence of many peaks
  gives a scaling dimension different both from disordered or periodic systems.}
\end{figure}

\begin{figure}[h]
\centerline{\psfig{file=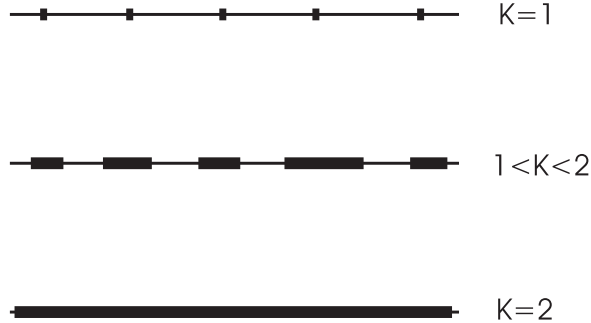,angle=0,width=\figwidth}}
\caption{\label{fig:spectrum} A possible evolution of the density
of states
 as a function of the Luttinger liquid parameter $K$. For $K=1$ the spectrum is a set
of measure zero and gaps are present almost everywhere. For $1<K<2$ the small gaps
close as $K$ increases. For $K=2$ even the largest gaps are closed
since even a periodic potential is irrelevant above $K=2$.}
\end{figure}

A more interesting behaviour is encountered when the Fermi level
is far from a dominant harmonic of the quasiperiodic potential,
for example at half-filling ($2k_F={\pi/a}$). Indeed, the low
energy properties up to $l_{max}$ are no more dominated by the
ultimate presence of a gap or not but by the precise dependence of
$G$ with the scale. As can be seen from figure~\ref{fig:smallpeak}
$G$ has an exponential scaling $G\sim e^{(2-2K)}l$.
Such a behaviour has several consequences. First it provides a
critical value of $K=1$ for the metal-insulator transition.
Let us point out that this value  is the smallest critical
value that one can expect for such a transition. Indeed it is
clear that repulsive interactions ($K<1$) enforce the insulating
character of the system. To know whether this analysis is
asymptotically correct or not, one would clearly need an
analytical calculation of $G$, a rather complicated task. Note
however that our predictions seem to be confirmed by numerical
work based on the Density Matrix Renormalization Group method on
the XXZ chain with quasiperiodic exchange \cite{Hida2}. This work
provides evidence for different behaviours for the scaling of the
gap with the size for various $K$. In particular, a transition is
observed for $K=1$ at half-filling even in the strong coupling
regime. An interesting, and yet unanswered question is how the
spectrum, or more generally the density of states is modified when
$K$ varies. It is very natural to assume that the small gaps close
up when $K$ increases, given that the system is always gapless for
$K=2$. The spectrum would thus evolve from a set of zero measure
for $K=1$ to a set of finite measure for $K > 1$ as indicated in
Figure~\ref{fig:spectrum}. It would be interesting to check this
scenario on the XXZ chains under a magnetic field.
So, the quasicrystal differs from a periodic one, for which the
gap only acts for a given position of the Fermi level with
$K_c=2$, and a disordered system for which the potential is
relevant regardless of the position of the Fermi level, but below
a constant critical value $K_c=3/2$. This important modification
of $K_c$ is reminiscent of a correlated disorder with long range
correlations in space for which the averaged disorder potential
$\overline{\hat W^*(q)\hat W(q')}=\delta_{qq'}\Delta(q)$ is not
constant.

The other important consequences of such a behaviour are of course
exhibited by the transport properties. A straightforward
calculation would give a power law for the frequency dependence of
the conductivity $\sigma(\omega) \sim (1/\omega)^{5-2K - \mu}$
with $\mu = 2$ (at variance with $\mu=0$ for periodic and $\mu=1$
for disordered). This anomalous power law depending {\bf both} on
the interactions and the spectrum properties would also appear in
a size dependence of the Landauer conductance $R(L) \sim
L^{4-2K - \mu}$. The temperature dependence of the conductivity
raises an interesting question, similar to the one occuring for
periodic systems. Cutting the flow at a the natural lengthscale
$l=\log(v_F/T)$ would lead to a power law dependence of the
conductivity $\sigma(T) \sim T^{3-2K-\mu}$, and thus to a finite
conductivity. In order to obtain such a behaviour it is probably
necessary to introduce dephasing processes (like a coupling to a
thermal bath). If such a temperature dependence of the
conductivity occurs it would be again a demonstration of the close
connection between quasiperiodic systems and disordered one (for
which it is quite natural to expect dissipation at finite
temperature).

Let us note that is it possible to investigate the effects of
disorder on the quasiperiodic potential using (\ref{eq:rgeq}).
Disorder in $d=1$ leads in itself to localization, which is to be
avoided if one wants to make connection with higher dimensions.
One way to achieve this is simply to introduce forward scattering
on impurities which cannot lead to localization
\cite{giamarchi_loc}. Introducing such a disorder cuts the flow in
(\ref{eq:rgeq}) at a characteristic lengthscale $\xi_F$ inversely
proportional to the disorder. Since it prevents the
(quasi-)periodic potential to grow it makes thus the system {\bf
more} conducting (more details can be found in
\cite{vidal_quasiinter_long}). One thus recovers on this simple
model the physical effects leading to the inverse Mathiessen law
experimentally observed.

Exploring these issues, as well as investigating the consequences
in higher dimension is a challenging problem left for the future.


\begin{thebibliography}{10}

\bibitem{Shechtman}
D.~Shechtman, I.~Blech, D.~Gratias, and J.~W. Cahn.
 {\em Phys. Rev. Lett.}, 53:1951, 1984.

\bibitem{Mayou93}
D.~Mayou, C.~Berger, F.~Cyrot-Lackmann, T.~Klein, and P.Lanco.
 {\em Phys. Rev. Lett.}, 70:3915, 1993.

\bibitem{Klein}
T.~Klein, C.~Berger, D.~Mayou, and F.~Cyrot-Lackmann.
 {\em Phys. Rev. Lett.}, 66:2907, 1991.

\bibitem{Biggs}
B.~D. Biggs, S.~J. Poon, and N.~Munirathan.
 {\em Phys. Rev. Lett.}, 65:2700, 1990.

\bibitem{Delahaye}
J.~Delahaye, J.~P. Brison, and C.~Berger.
 {\em Phys. Rev. Lett.}, 81:4204, 1998.

\bibitem{Homes}
C.~Homes, T.~Timusk, X.~Wu, Z.~Altounian, A.~Sahnoune, and J.~O.
  Str{\"o}m-Olsen.
 {\em Phys. Rev. Lett.}, 67:2694, 1991.

\bibitem{Burkov}
S.~Burkov, T.~Timusk, and N.~W. Aschroft.
 {\em J. Phys. C}, 4:9447, 1992.

\bibitem{KKT}
M.~Kohmoto, L.~P. Kadanoff, and C.~Tang.
 {\em Phys. Rev. Lett.}, 50:1870, 1983.

\bibitem{Ostlund}
S.~Ostlund, R.~Pandit, D.~Rand, H.~J. Schellnhuber, and E.~D. Siggia.
 {\em Phys. Rev. Lett.}, 50:1873, 1983.

\bibitem{Zhong}
J.~X. Zhong and R.~Mosseri.
 {\em J. Phys. C}, 7:8383, 1995.

\bibitem{Piechon}
F.~Pi{\'e}chon.
 {\em Phys. Rev. Lett.}, 76:4375, 1996.

\bibitem{Passaro_Octo_diffusion}
B.~Passaro, C.~Sire, and V.~G. Benza.
 {\em Phys. Rev. B}, 46:13751, 1992.

\bibitem{Yamamoto_Penrose}
S.~Yamamoto and T.~Fujiwara.
 {\em Phys. Rev. B}, 51:8841, 1995.

\bibitem{Mayou99}
D. Mayou, 1999, cond-mat 9912113.

\bibitem{Sire_Aussois}
C.~Sire.
 {\em Lectures on Quasicrystals}.
 Editions de Physique, Les Ulis France, 1994.

\bibitem{Roche_Review}
S.~Roche, D.~Mayou, and G.~{Trambly de Laissardiere}.
 {\em J. Math. Phys.}, 38:1794, 1997.

\bibitem{Roche_Fermi}
S.~Roche and T.~Fujiwara.
 {\em Phys. Rev. B}, 58:11338, 1998.

\bibitem{Bellissard_aperiodic}
H.~Schulz-Baldes and J.~Bellissard.
 {\em J. Stat. Phys.}, 91:991, 1998.

\bibitem{Tsunetsugu_conductance}
H.~Tsunetsugu and K.~Ueda.
 {\em Phys. Rev. B}, 43:8892, 1991.

\bibitem{Kubo_Goda}
Masaki Goda and Haruhiko Kubo.
 {\em J. Phys. Soc. Jpn.}, 58:2109, 1989.

\bibitem{Hiramoto_HF}
H.~Hiramoto.
 {\em J. Phys. Soc. Jpn.}, 59:811, 1990.

\bibitem{Chaves_meso}
J.~C. Chaves and I.~I. Satija.
 {\em Phys. Rev. B}, 55:14076, 1997.

\bibitem{Chaves_HF}
J.~C. Chaves and I.~I. Satija, 1998.
 cond-mat 9803103.

\bibitem{Mastro}
Vieri Mastropietro, 1998.
 cond-mat 9810128.

\bibitem{Sen}
D.~Sen and S.~Lal, 1998.
 cond-mat 9811330.

\bibitem{vidal_boso}
J.~Vidal, D.~Mouhanna, and T.~Giamarchi.
 {\em Phys. Rev. Lett.}, 83:3908, 1999.

\bibitem{schulz_houches_revue}
H.~J. Schulz.
 Elsevier, Amsterdam, 1995.

\bibitem{haldane_bosonisation}
F.~D.~M. Haldane.
 {\em J. Phys. C}, 14:2585, 1981.

\bibitem{haldane_xxzchain}
F.~D.~M. Haldane.
 {\em Phys. Rev. Lett.}, 45:1358, 1980.

\bibitem{vidal_quasiinter_long}
J. Vidal et al., to be published.

\bibitem{luther_exact}
A.~Luther and V.~J. Emery.
 {\em Phys. Rev. Lett.}, 33:589, 1974.

\bibitem{giamarchi_umklapp_1d}
T.~Giamarchi.
 {\em Phys. Rev. B}, 44:2905, 1991.

\bibitem{kolomeisky_periodic_diagphas}
E.~Kolomeisky.
 {\em Phys. Rev. B}, 47:6193, 1993.

\bibitem{footnote_harper}
This point was missed in previous studies of a quasiperiodic potential, that
  considered a Harper like potential $\cos(Q x)$
  \cite{kolomeisky_periodic_diagphas}. Indeed, as was already known for the
  noninteracting case, such a potential is perturbatively very similar to a
  simple periodic one, and quasiperiodic effects only occur markedly for large
  enough potential, making it unsuitable for an RG study. The Fibonacci
  potential studied here does not suffer from such problems and exhibit the
  quasiperiodic effects even at small coupling.

\bibitem{giamarchi_loc}
T.~Giamarchi and H.~J. Schulz.
 {\em Phys. Rev. B}, 37:325, 1988.

\bibitem{apel}
W.~Apel.
 {\em J. Phys. C}, 15:1973, 1982.

\bibitem{Mosseri_conumbering}
R.~Mosseri.
 In J.~Yacam{\'a}n {\it et al.}, editor, {\em Proceedings of the 3rd
  International Meeting on Quasicrystals}, page 129, Singapore, 1990. World
  Scientific.

\bibitem{Sire_pertu}
C.~Sire and R.~Mosseri.
 {\em J. Phys. (Paris)}, 50:3447, 1989.

\bibitem{Hida2}
K.~Hida, 1999.
 cond-mat~ 9911424.

\end{thebibliography}

\end{document}